\documentclass[12pt]{article}

\usepackage{amssymb,amsmath,}
\usepackage{amstext}
\usepackage{amscd}
\usepackage{slashed}
\usepackage{graphics}
\usepackage{epsfig}
\def\wt{\widetilde}
\newcommand{\Rho}{{\mbox{\sf P}}}

\def\Real{\mathbb R}

\def\ts{\textstyle}

\def\be{\begin{equation}}
\def\ee{\end{equation}}
\def\beq{\begin{equation}}
\def\eeq{\end{equation}}
\def\bea{\begin{eqnarray}}
\def\eea{\end{eqnarray}} 

\def\eqn#1{(\ref{#1})}

\def\nn{\nonumber}

\def\sideremark#1{\ifvmode\leavevmode\fi\vadjust{\vbox to0pt{\vss
 \hbox to 0pt{\hskip\hsize\hskip1em
 \vbox{\hsize3cm\tiny\raggedright\pretolerance10000
  \noindent #1\hfill}\hss}\vbox to8pt{\vfil}\vss}}}

\begin{document}

\thispagestyle{empty}

\vspace{.8cm}
\setcounter{footnote}{0}
\begin{center}
{\Large
 {\bf 
The $\frak{so}(d+2,2)$ Minimal Representation and  Ambient Tractors:
 the Conformal Geometry of Momentum Space.}\\[10mm]


 {\sc \small A.R.~Gover$^{\mathfrak G}$ and A.~Waldron$^{\mathfrak W}$\\[6mm]}
 {\em\small 
            Department of Mathematics 
            }\\[6mm]
 {\em\small ${}^{\mathfrak G}\!$
            The University of Auckland,
            Auckland PB 1142, New Zealand\\
            {\tt gover@math.auckland.ac.nz}\\[5mm]      
            ${}^{\mathfrak W}\!$
            University of California,
            Davis CA 95616, USA\\[-3mm] 
            {\tt wally@math.ucdavis.edu}
            }\\[5mm]

}

\bigskip

\bigskip

{\sc Abstract}\\
\end{center}

{\small
\begin{quote}

Tractor Calculus is a powerful tool for analyzing Weyl invariance;
although fundamentally linked to the Cartan connection, it may also be  
arrived at geometrically by viewing a conformal manifold as the space of
null rays in a Lorentzian ambient space.  For dimension $d$
conformally flat manifolds we show that the $(d+2)$-dimensional
Fefferman--Graham ambient space corresponds to the momentum space of a
massless scalar field. Hence on the one hand the null cone
parameterizes physical momentum excitations, while on the other hand,
null rays correspond to points in the underlying conformal
manifold. This allows us to identify a fundamental set of tractor
operators with the generators of conformal symmetries of a scalar
field theory in a momentum representation.  Moreover, these constitute
the minimal representation of the non-compact conformal Lie symmetry
algebra of the scalar field with Kostant--Kirillov dimension
$d+1$. Relaxing the conformally flat requirement, we find that while
the conformal Lie algebra of tractor operators is deformed by
curvature corrections, higher relations in the enveloping algebra
corresponding to the minimal representation persist. We also discuss
potential applications of these results to physics and conformal
geometry.

\bigskip

\bigskip

\end{quote}
}

\newpage

\section{Introduction}

The massless wave equation is ubiquitous in both Physics and
Mathematics.  In this Article we point out yet another {\it r\^ole}
for this equation in spaces of metric signature $(p-1,q-1)$: (in a
suitably flat setting) its solution space describes the conformal
geometry of pseudo-Riemannian manifolds of signature $(p-2,q-2)$. The
key idea is that solutions to the flat space wave equations on
$\Real^{p-1,q-1}$ are parameterized by lightlike momenta. In this
picture the space of all possible ``off-shell'' momenta corresponds to
the ambient space of a $d=(p+q-4)$-dimensional conformal manifold; the
latter can be viewed as the space of lightlike rays in a certain null
hypersurface.  Although our results hold in any signature 
$p,q>1$ (with $p+q\geq 5$), we set $(p,q)=(d+2,2)$ for simplicity of exposition; we leave
it as an exercise for the reader to make the minor adjustments to
recover other signatures. We will call the space~$\Real^{d+1,1}$
hosting the scalar wave equation the ``dual ambient space''.

The duality that we study in this paper is defined via Fourier
transform. An important feature of this construction are the
symmetries following from the dual ambient conformal group
$SO(d+2,2)$. It is well known that solutions to the scalar wave
equation form an irreducible unitary representation of this group.  In
fact, a recent study by Kobayashi and $\slashed{\rm O}$rsted~\cite{KO}
gives an explicit construction of the intertwiner by Fourier transform
of the reducible $SO(d+2,2)$ representation on the dual ambient space.
The  restriction of which recovers an
irreducible one on the ambient lightcone and they identify it (for
$d\geq4$) as the minimal representation (see~\cite{K,BZ}).  This idea
of studying conformal symmetry by looking at spaces of two higher
dimensions dates back to Dirac~\cite{Dirac}, but the additional Fourier
duality gives a great deal of geometric information.

In this Article, the differential generators of ${\frak{so}}(d+2,2)$ acting on
the dual ambient space will be shown to play a distinguished geometric
{\it r\^ole} once intertwined to act on the ambient lightcone. The
relevant geometry is described by the tractor calculus for conformal
manifolds of~\cite{Thomas,BEG}. Central to the tractor calculus, is a
basic set of differential operators that were constructed at first in
an apparently {\it ad hoc} way in~\cite{Thomas,BEG,Gsrni,G}.  In fact,
these are strongly linked to representation theory and collectively
arise from the fundamental derivative found in~\cite{CapGoTAMS}; the
latter is a first order differential operator with universal
properties, existing on a large range of structures.  Later, these
tractor operators were recovered~\cite{CG,GP} from the
Fefferman--Graham (FG) ambient manifold~\cite{FG} associated
canonically with any conformal manifold (with dimension greater than
two). Here we show that, at least in the conformally flat case, the
collection of tractors operators is unified and arises independently
from the picture above. In fact they
are the differential generators of the minimal representation on the
lightcone.  Therefore, from this point of view, their origin is as the generators of the {\it
  dual} ambient conformal group. This relationship is expressed by the
intertwiner by Fourier transform. It follows that, in the conformally flat setting, 
the tractor operators (see~\eqn{fgens}) obey an ${\frak{so}}(d+2,2)$ algebra (see~\eqn{lie algebra}). Moreover, the underlying minimal representation
carries further geometric information in the form of quadratic
relations among tractor operators. Returning to generally curved
conformal manifolds, the ${\frak{so}}(d+2,2)$ Lie algebra structure is deformed
to reflect the curvature of the ambient manifold. Remarkably, we show
that these quadratic relations continue to hold, and encode the key
identities obeyed by the tractor operators themselves. Although these
identities were known previously, this new derivation provides a
structure explaining why these identities exist and suggests an
obvious extension to other geometric structures.

The remainder of this Article is organized as follows. We begin by giving some key facts about tractors and review the
FG ambient metric construction for a  conformal manifold in Sections~\ref{Tractors},~\ref{ambient}.
Then we introduce the massless scalar field theory and its conformal symmetries in
Section~\ref{scalar}. In Section~\ref{tractors} we establish the relationship between tractor
operators and the ambient conformal symmetries. In Section~\ref{curved} we
show how to generalize our results to curved ambient spaces and conformal manifolds that
are not conformally flat. Potential applications to both physics and conformal geometry are
discussed in the Conclusions.

\section{Tractors}

\label{Tractors}

{\em Weyl transformations} \be g_{\mu\nu} \mapsto
\widehat{g}_{\mu\nu}=\Omega^{2} g_{\mu\nu}\, ,\label{Weyl} \ee are fundamental
to both physics and mathematics. In particular, conformal geometry is
the study of manifolds equipped with a (conformal) equivalence class
of metrics where the equivalence relation is given by $g\sim
\widehat{g}$ if $g$ and $\widehat{g}$ are related as in (\ref{Weyl}),
and where $\Omega$ is any non-vanishing function on the given
$d$-dimensional manifold $M$.  

In pseudo-Riemannian geometry the standard invariant calculus is based
around the Levi Civita connection $\nabla_\mu$ determined canonically
by the metric~$g_{\mu\nu}$.  The fundamental local invariant is the
Riemannian curvature $R_{\mu\nu}{}^m{}_n$ and we recall that this
decomposes according to 
$$
R_{\mu\nu\rho\sigma}-W_{\mu\nu\rho\sigma}=\Rho_{\mu\rho}g_{\nu\sigma}
-\Rho_{\nu\rho}g_{\mu\sigma}
-\Rho_{\mu\sigma}g_{\nu\rho}
+\Rho_{\nu\sigma}g_{\mu\rho}\, ,
$$
where $W$ is the completely trace-free part of the curvature $R$ and
$\Rho$ is the Schouten tensor.  Descending to the equivalence class of
metrics of conformal geometry there is no satisfactory
conformally invariant connection on the tangent bundle and this
frustrates na\"ive attempts to construct well defined natural
equations and invariants. On the other hand, in dimensions $d\geq 3$
(and we make this restriction throughout) there is a conformally
invariant connection on a bundle of rank $d+2$ which extends the
tangent bundle. In a metric scale this (standard) {\em tractor} bundle, denoted
${\cal E}^M$ (in a Penrose notation), is simply a direct sum of the
tangent bundle with two copies of the trivial bundle and the {\em
  tractor connection} is given explicitly by \be {\cal D}_\mu
\left(\begin{array}{c}
    T^+\\[2mm]
    T^m\\[2mm]
    T^-
\end{array}
\right)
=
\left(\begin{array}{c}
\nabla_\mu T^+ - T_\mu\\[2mm]
\nabla_\mu T^m+\Rho_\mu{}^m T^+ + e_\mu{}^m T^- \\[2mm]
\nabla_\mu T^- - \Rho_{\mu n} T^n 
\end{array}
\right)\, ,\label{vector connection} \ee 
where $T^+$ and $T^-$ are
functions while $T^m$ is a tangent vector field.  This connection is
equivalent to the normal Cartan connection of \cite{Cartan}, see
\cite{CapGoTAMS}, but the first construction and use of it as here was 
(independently) by Thomas~\cite{Thomas}.  

The connection ${\cal D}_\mu$ preserves a conformally invariant
tractor metric $h_{MN}$ (of signature $(d+1,1)$ for Riemannian conformal geometry) which is given as a quadratric
form on $T^M$ by the formula $h_{MN}T^MT^N=2T^+T^- + T^m T_m$. Thus we
see some analogies with the pseudo-Riemannian calculus. However
observe that the tractor covariant derivative of $T^M$, {\it viz}.\ $D_\mu
T^M$ is a mixed 1-form--tractor and so the route to higher
derivatives, captured in some conformally invariant way, is not
immediately clear. Solving this in general turns out to be rather
subtle and this is one of the key applications of tractor calculus.
This point of view has its origins in \cite{Thomas} and was
rediscovered and developed significantly in
\cite{BEG,G,CapGoTAMS,GP,CG}.  It first appeared in a physics context in the supergravity-motivated 
study of conformal gravity~\cite{Kaku:1978nz} and these ideas have also been applied
recently to show how Weyl invariance underlies the origins of mass in
particle equations \cite{Gover:2008sw,Gover:2008pt}.  Among these
\cite{BEG,GP,CG,Gover:2008sw}, for example, give background for details of the
calculations described. 

Part of the solution, to the alluded problem, involves a collection of
conformally invariant differential operators that can be iterated
without moving out of the tractor framework and these are the subject
of the current Article. Before we describe these, some notation is required.
We will often work with tractor bundles with  (or ``twisted by'') a conformal weight which may be denoted $w$. We write  ${\cal E}^M[w]$ for the  
weight $w$ tractor bundle.
Weighted tensor powers of
the standard tractor bundle are denoted ${\cal E}^{MN\ldots}[w]$ and
accordingly with $(\cdots)$ and $[\cdots]$ denoting symmetrization and
antisymmetrization, respectively.  
So with this notation the
key tractor-valued operators of our focus act as follows:
\bea
X^M \ :{\cal E}^{N\ldots}[w]& \rightarrow &{\cal E}^{MN\ldots}[w+1]\, ,\nn\\[4mm]
D^{[MN]}:{\cal E}^{R\ldots}[w]&\rightarrow&{\cal E}^{[MN]R\ldots}[w]\, ,\nn\\[4mm]
D^M  \ :{\cal E}^{N\ldots}[w]& \rightarrow &{\cal E}^{MN\ldots}[w-1]\, .
\label{operators}
\eea
The first of these is defined by multiplication by the canonical tractor $X^M$. The Thomas-$D$-operator $D^M$ and double-$D$ operator $D^{MN}$ are defined in the standard way. To this set of tractor operators, it will be important to add the operator whose eigenvectors are tractors of definite weight 
and eigenvalue is the weight. The operators in~\eqn{operators}  can also be expressed explicitly in terms of the tractor connection for a given choice of metric as  
\be
X^M \! =\! \left(\begin{array}{c}0\\0\\1\end{array}\right)  ,\quad\!\!
D^{[MN]}\!=\!
\left(\!\!
\begin{array}{ccc}
0&0&w\, \\[1mm]0&0&{\cal D}^m\\[1mm] -w &-{\cal D}^n&0
\end{array}
\!\!\right) ,\quad\!\!
D^M\!  = \!\left(\!\!\begin{array}{c}w(d+2w-2)\\[1mm](d+2w-2){\cal D}^m\\[1mm]-({\cal D}^r {\cal D}_r+w\Rho)\end{array}\!\!\right) .
\ee
That they map objects covariant under tractor gauge transformations to objects with the
same covariance can be checked by explicit calculation but also follows from general
principles~\cite{BEG}. More importantly, their extensive practical applicability to constructing Weyl invariants
and covariants, as well as physical theories from first principles, is clear~\cite{G,GP,BGo,Gover:2008sw,Gover:2008pt}.

\section{The Conformal Ambient Metric Construction}

\label{ambient}

\begin{figure}
\begin{center}
\epsfig{file=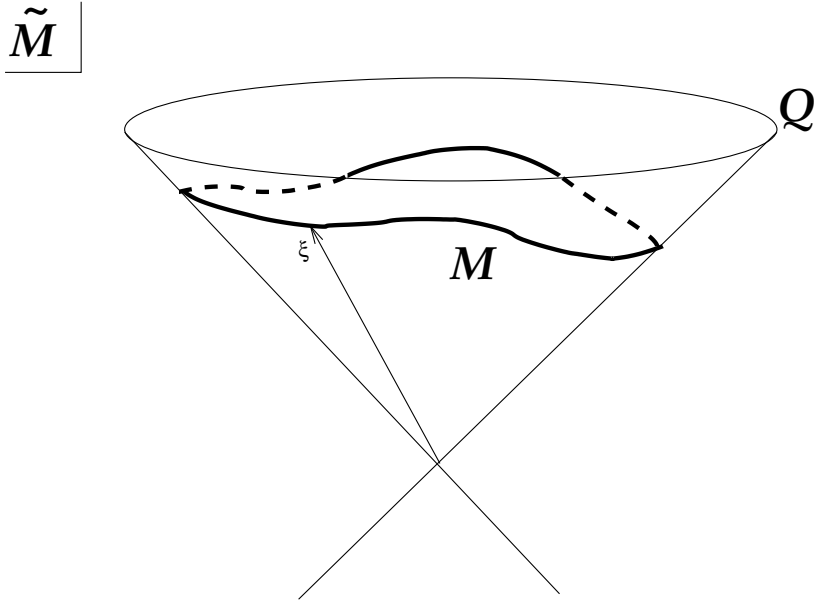,height=8cm}
\end{center}
\caption{The sphere realized as rays $\xi$ of the cone $Q$
in the ambient space~$\wt M$.\label{cone}}
\end{figure}

\begin{figure}
\begin{center}
\epsfig{file=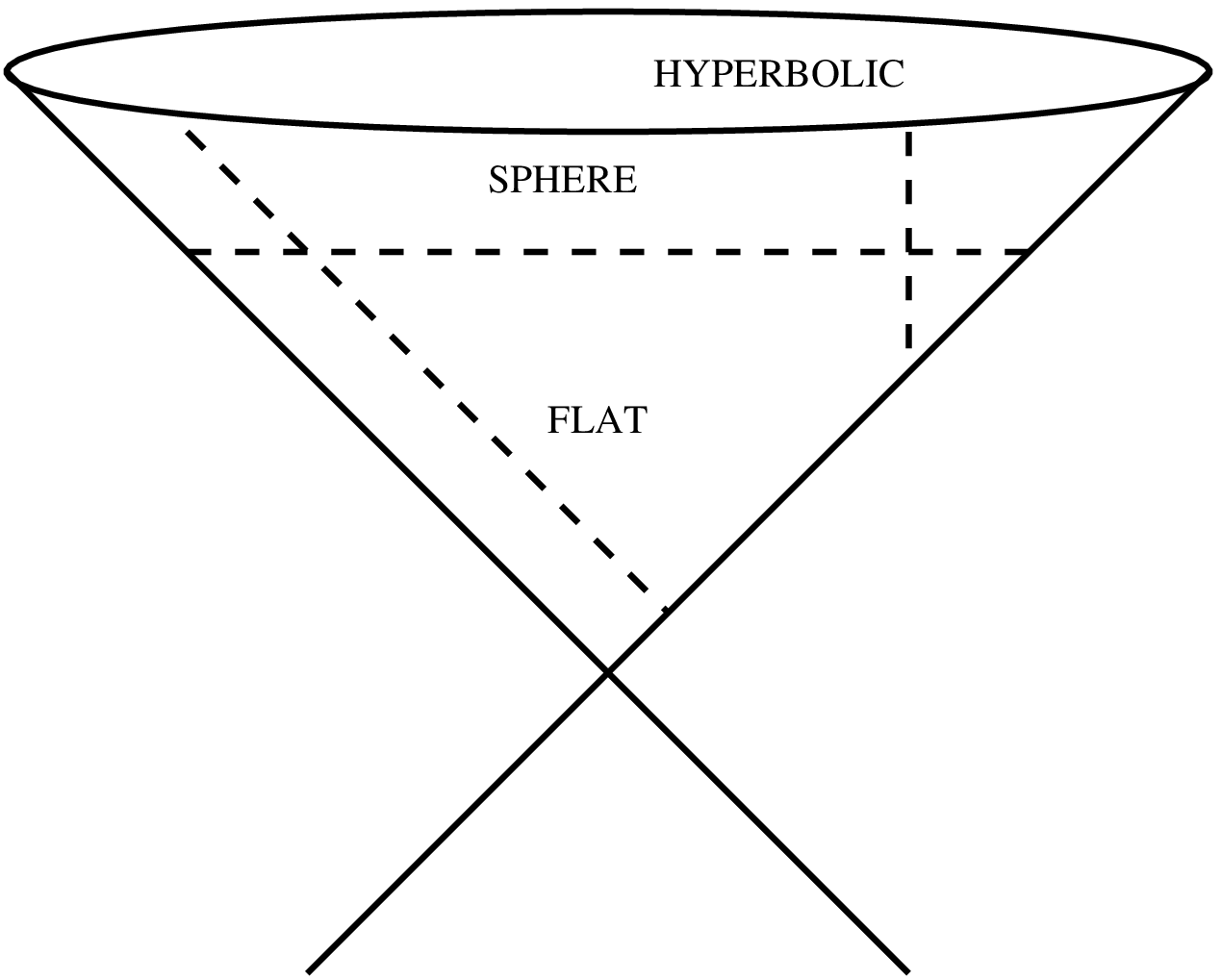,height=8cm}
\end{center}
\caption{The sphere, flat space and hyperbolic space obtained from  slices of the cone $Q$.\label{slices}}
\end{figure}

A useful way to view a $d$-dimensional conformal manifold $(M,[g_{\mu\nu}])$
is as the space of rays in a $(d+1)$-dimensional null hypersurface $Q$ in a $(d+2)$-dimensional
Riemannian ambient space $(\wt M,h_{MN})$. More precisely---special\-izing to the conformally flat case---consider the ambient space $\wt M=
\Real^{d+1,1}$ with the standard flat Lorentzian metric
\be
\wt{ds}{}^2=dX^M \eta_{MN} dX^N\, . \nn
\ee
This metric enjoys a hypersurface orthogonal homothety given by the Euler operator\footnote{We use $\pounds$ to denote the Lie derivative.}
\be
{\bf w}=X^M\frac{\partial}{\partial X^M}\, ,\qquad
\pounds_{\bf w} \wt{ds}^2 = 2 \wt{ds}^2\, .
\ee
The zero locus of the homothetic potential
\be
H=\frac12X^M X_M\equiv \frac 12\,  X^2\, , \nn
\nn\ee
defines the null cone $Q$. $H$ is the defining function for the cone $Q$. The space of null rays in $Q$, {\it viz.}, solutions
$\xi^M$ to $H=0$ subject to the equivalence relation
\be
\xi^M\sim \Omega\,  \xi^M\, ,\qquad \Omega\in \Real^+\, ,
\nn\ee
is $(d+1)$-dimensional and may be identified with the conformal manifold $M$.
This is depicted in Figure~\ref{cone}.
We may obtain the conformal class of metrics as follows:  Let 
$\xi^M(x)$ denote a section of the null cone  $Q$.
The ambient metric pulls back to a metric $ds^2=d \xi^M d\xi_M$  on the submanifold. 
Identifying this submanifold with $M$ yields a metric
$ds^2$ on $M$. Choosing a different section results in a conformally related metric.
For example, in the conformally flat setting, the sphere,
flat  and hyperbolic metrics all inhabit the same conformal class. In the ambient picture
they correspond to slicing the cone horizontally, diagonally and vertically, respectively. This 
is depicted in Figure~\ref{slices}.

We will obtain the tractor operators in this ambient picture by reinterpreting the 
null cone~$Q$ as the space of massless excitations for a scalar field. In more mundane
language, we propose interpreting $Q$ as the onshell lightcone for a scalar field. In particular, in this picture,
the ambient {\it coordinates}~$X^M$ must be viewed as {\it momenta}. Before studying scalar fields,
however, let us briefly indicate the ambient origin of the tractor connection~${\cal D}_\mu$:
Firstly observe that the Lorentz group $G=SO_o(d+1,1)$ acts naturally on the ambient space $\wt M=\Real^{d+1,1}$. Then pick any future lightlike ray $\xi$ in $\wt M$ and call its stabilizer $P$. The
coset $G/P$ is exactly the sphere $M$. Indeed, this is the classical model for the sphere
equipped with a conformally flat structure. The Maurer--Cartan form on $G$ is ${\frak{so}}(d+1,1)$-valued
so when pulled back from~$G$ to $M=G/P$, it defines a ${\frak{so}}(d+1,1)$ Yang--Mills connection. 
This connection is precisely the tractor connection. Its equivalent parallel transport is that arising from
ambient pseudo-Euclidean parallel transport.

\section{Ambient Scalar Fields}

\label{scalar}

Unitary representations for non-compact forms of orthogonal groups are
necessarily infinite dimensional. There is however a notion of
Gel'fand--Kirillov dimension~\cite{BZ,KO} associated with the number
of independent variables for unitary representations on function
spaces and in turn representations that are in this sense
minimal~\cite{Joseph}.  For orthogonal groups $O(p,q)$, the minimal
unitary representation is shown to be related to the space of zero
modes of the ultrahyperbolic Laplace operator on $\mathbb
R^{p-1,q-1}$~\cite{BZ,KO}.  This setting is of course a familiar one
in physics, namely a free massless scalar field in a flat
background---essentially the simplest quantum field theory possible.
Here we sketch physical aspects of this construction of the minimal
unitary representation of $O(d+2,2)$ in order to develop the intuition
needed for our construction of a symmetry algebra for tractor
operators. (We have chosen this particular signature so that our
underlying conformal manifold will be Riemannian. Our results extend
immediately to conformal manifolds of any metric signature, however.)
Consider therefore a scalar field $\Phi(Y)$ in flat $\Real^{d+1,1}$
with local coordinates $Y^M$. For reasons which will become clear, we
call this space the {\it dual ambient space}. Let the dynamics of this
scalar field be governed by the action principle \be S=-\frac12 \int
d^{d+2}Y \ \frac{\partial \Phi}{\partial Y^M} \ \eta^{MN}\,
\frac{\partial \Phi}{\partial Y^N}\, .
\label{S}
\ee
This action is   invariant under
${\frak{so}}(d+2,2)$ rigid symmetry transformations corresponding to the conformal Killing vectors
\bea
&iP_M\equiv\frac{\ts \partial}{\ts \partial Y^M} \ \ \mbox{  (translations)}&\nn\\[2mm]
D\equiv Y^M\frac\partial{\partial Y^M}\ \  \mbox{  (dilations)}
\hspace{-2.7cm}&&\hspace{-6.5cm}
M_{MN}=Y_M\frac\partial{\partial Y^N}-Y_N\frac\partial{\partial Y^M}\ \ \mbox{ (rotations)}\nn\\[2mm]
&iK_M=-2Y_M D+Y^NY_N \  \frac{\ts \partial}{\ts \partial Y^M}
\ \ \mbox{  (conformal boosts)}
\, .&\label{ckilling}
\eea
Invariance of the action~\eqn{S} is achieved by transformations of $\Phi$ with a conformal weight $-d/2$
generated by
\be
\delta_X \Phi = \pounds_X \Phi + \frac d2\, \alpha_X  \Phi\, ,\label{symmetries}
\ee
where $X$ stands for any of the vectors~\eqn{ckilling} and the function $\alpha_X=(d+2)^{-1}{\rm div} X$ so that $\pounds_X \wt ds^2 = 2 \alpha_X \wt ds^2$. In particular $\alpha_D=1$ and $\alpha_{K_M}=
2iY_M$. 

The operators $\delta_X$ in~\eqn{symmetries} acting on scalars generate a
(reducible) representation of ${\frak{so}}(d+2,2)$. Explicitly the non-vanishing Lie brackets of the above vector fields are
$$
[M_{MN},M_{RS}]=\eta_{NR}M_{MS}-\eta_{MR}M_{NS}-\eta_{NS}M_{MR}+\eta_{MS}M_{NR}\, ,
$$
\vspace{-.8cm}
\bea
[D,P_M]=-P_M\, ,&& [D,K_M]=K_M\, ,\nn\\[3mm]
[M_{MN},P_R]=2P_{[M}\eta_{N]R}\, ,&&[M_{MN},K_R]=2K_{[M}\eta_{N]R}\, ,\nn
\eea
\be
[P_M,K_N]=2\eta_{MN}D-2M_{MN}\, .\label{lie algebra}
\ee
 Our aim is to relate the operators $\pounds_X+\frac d2 \alpha_X$  to tractor operators.
This involves identifying the generators of the conformal isometry group $SO(d+2,2)$ of the dual ambient space with
tractor operators on an ambient space~$\wt M$.
However, it is important to already stress that the $\Real^{d+1,1}$ dual ambient space on which our scalar field
theory lives is {\it not} the ambient manifold~$\wt M$ but rather will be related to it by a Fourier transformation. To study this relationship we first solve the classical field theory defined by~\eqn{S}.

The field equation implied by the action~\eqn{S} \be\Delta \Phi=0\,
,\label{eom}\ee may be solved by Fourier transform \be \Phi(Y)=\int
\frac{d^{d+2}X\ \delta(X^2)}{(2\pi)^{d+2}} \ \exp(iX_M Y^M)\,
\varphi(X)\, ,\label{intertwine} \ee whose integrand has delta
function support on the light cone $Q$. This mapping between solutions
to the ultrahyperbolic wave equation~\eqn{eom} and distributions
supported on the null cone $Q$ with defining function $H=\frac12 X^2$
is constructed in detail in~\cite{KO}.  Since our main focus is
identifying the image of the conformal ambient ${\frak{so}}(d+2,2)$
differential generators under the Fourier
intertwiner~\eqn{intertwine}, we shall avoid a detailed discussion of
the function spaces involved.  Physically the set of variables $(X_M)$
is interpreted as an $\Real^{d+1,1}$-valued momentum.  We advocate
identifying this momentum space with the ambient manifold $\wt M$.  By
including the $\delta$ distribution explicitly we execute a truncation
of $\varphi (X)$. Even though only values of $\varphi(X)$ along the cone are
important for the integral, to recover calculus with ambient content
we shall impose the condition that smooth $\varphi(X)$ on the cone are
also extended off smoothly; that is we take $\varphi(X)$ to be smooth on the
ambient space.
In particular the solution space may be
viewed as an equivalence class of functions \be \varphi(X) \sim
\varphi(X) + H \xi(X)\, ,\label{equivalence} \ee for a suitable class
of functions $\xi(X)$. For our purposes, smooth functions suffice.

The Fourier transform~\eqn{intertwine} is an intertwiner between
${\frak{so}}(d+2,2)$ representations. Acting on arbitrary functions of $\wt M$
(regardless of their support away from $Q$) it is simple to compute
the differential representation of ${\frak{so}}(d+2,2)$ using the usual
correspondence under Fourier transforms $Y^M\rightarrow
i\partial\slash \partial X_M$ and $\partial\slash \partial
Y^M\rightarrow i X^M$.  The resulting operators are not vector fields
on $\wt M$ (save for the rotations $M_{MN}$) but, necessarily, obey
the ${\frak{so}}(d+2,2)$ Lie algebra. Also, they are not quite the operators we
need. Instead we are interested in what becomes of the
symmetry operators $\pounds_X+\frac d2 \alpha_X$ when passed through
the intertwiner~\eqn{intertwine}, including the delta
function~$\delta(X^2)$, so that they act on the physical
moduli~$\varphi(X)$. In this simple flat setting, this computation can
be done explicitly by actually performing the delta function
integration. To contrast them from their underlying Killing
vectors~\eqn{ckilling}, we denote these intertwined symmetry
generators in bold face and find the following operators agree with the action of the conformal Killing symmetry operators
\bea
&{\bf P}_M=X_M  &\nn\\[2mm]
{\bf D}=-X^M\frac\partial{\partial X^M}-d/2
\hspace{-1.0cm}&&\hspace{-2.1cm}
{\bf M}_{MN}=X_M\frac\partial{\partial X^N}-X_N\frac\partial{\partial X^M} 
\nn\\[3mm]
&{\bf K}_M=-2 {\bf D}\frac{\ts \partial}{\ts \partial X^M}
-X_M \frac{\ts \partial}{\ts\partial X^N}\frac{\ts \partial}{\ts \partial X_N}
\, .&
\label{fgens}
\eea An important consequence of the symmetries~\eqn{symmetries} is
that dual ambient space conformal transformations map solutions
of~\eqn{eom} to solutions. Working to the right of the Dirac
distribution (as a multiplication operator) for consitency with our
point of view above we require image operators that respect the
equivalence relation~\eqn{equivalence}. To see this explicitly for the
operators proposed, we compute the commutators of the
generators~\eqn{fgens} with the defining function~$H$, the
non-vanishing results are \be [{\bf D},H]=-2H\, ,\qquad [{\bf
    K_M},H]=4H\partial_M\, .\label{first class} \ee Hence if
$\varphi\sim\wt \varphi$ under~\eqn{equivalence}, then ${\bf X}
\varphi \sim {\bf X} \wt \varphi$ for ${\bf X}$ equaling any of the
generators~\eqn{fgens}. This was noted by~\cite{GP} where it was
proven that the operators~\eqn{fgens} act tangentially to the cone
defined by $H$. Here we see that this result enables natural ambient
extensions of the Fourier duals to the conformal symmetries of the
massless scalar field theory~\eqn{S}. Before recasting this physical
theory in terms of tractors and conformal geometry, we briefly discuss
the minimal representation and some of its consequences.
\subsection{The Minimal Representation}

Our computation began with off-shell massless
scalar fields $\Phi(Y)$---functions of $d+2$ variables---on which the group $SO(d+2,2)$ acts by conformal
transformations~\eqn{symmetries}. This is
not an irreducible representation but instead closely related to one. 
The Fourier transform~\eqn{intertwine} yields a representation on on-shell
massless fields in momentum space given by~\eqn{fgens}. 
As is well known in  a physical 
setting, this is a unitary and irreducible representation of $SO(d+2,2)$. 
Indeed, as shown in~\cite{KO}, this is the minimal representation of $SO(d+2,2)$ (for $d\geq 4$) and
the number of  variables required to coordinatize the cone $Q$ is its Kostant-Kirillov 
dimension $d+1$.  The Fourier transform is the intertwiner between these representations.

The minimal representation admits
a set of quadratic relations in the universal enveloping algebra~${\cal U}({\frak{so}}(d+2,2))$ which
 imply important tractor identities in a conformal geometry setting.  One may characterize representations by relations among
generators in the universal enveloping algebra. 
Consider, therefore, the matrix of ${\frak{so}}(d+2,2)$ symmetry generators
\be
J=\left(\begin{array}{ccc}
    -{\bf D}&-\frac1{\sqrt2}\ {\bf P}_M&0\\[3mm]
    \frac1{\sqrt2}\ {\bf K}_N&{\bf M}_{MN}&
                   \frac1{\sqrt2}\ {\bf P}_N\\[3mm]
    0&-\frac1{\sqrt2}\ {\bf K}_M&{\bf D}
        \end{array}\right)\, .
\nn\ee
In particular, 
notice that
\be
{\rm tr} J^2= {\bf M}_{MN}{\bf M}^{NM}+2{\bf D}^2
-{\bf P}^M{\bf K}_M-{\bf K}^M{\bf P}_M \, ,
\nn\ee
is the quadratic Casimir. In fact, for the representation~\eqn{fgens}
it obeys
\be
{\rm tr} J^2=-\frac12\ d(d+4)\, ,\label{c2}
\ee
a constant, as is necessarily the case for an irreducible representation.
However, we can search for even stronger relations by computing $J^2$ and
find, 
\be
(J+1)(J+d/2)=0\ \; {\rm mod}\; \ H\, ,
\ee
{\it I.e.}, the right hand side vanishes on the cone $Q$ in the sense of~\eqn{equivalence}.
Explicitly, this implies the following relations
\bea
&{\bf P}_M{\bf P}^M=0\, ,&\nn\\[2mm]
&{\bf P}_M({\bf D}+d/2)-{\bf P}^N {\bf M}_{NM}
=0\, ,&\nn\\[2mm]
&-\frac12\ {\bf P}_M{\bf K}_N+{\bf M}_M{}^R {\bf M}_{RN}
 -\frac12\ {\bf K}_M{\bf P}_N
 -(1+d/2)\ {\bf M}_{MN}+d/2\ \eta_{MN} =0\, , &\nn\\[2mm]
&({\bf D}+1)({\bf D}+d/2)-\frac12{\bf P}_N{\bf K}^N=0\, ,
  &\nn\\[2mm]
&({\bf D}+d/2){\bf K}_M- {\bf M}_{MN}{\bf K}^N=0\, ,&\nn\\[2mm]
&{\bf K}^M{\bf K}_M=0\, .&
\label{seven}
\eea
These relations all have tractor theoretic interpretations which we will analyze in the next Section.

\section{Tractors and the Ambient Conformal Group}

\label{tractors}

In~\cite{CG,GP} it is shown that the tractor operators~\eqn{operators}
may be recovered from (curved versions of) the ambient
operators~\eqn{fgens}. The latter act on ambient functions
$\varphi(X)$. However, if we impose both the equivalence
relation~\eqn{equivalence} and the homogeneity condition \be
X^M\frac{\partial}{\partial X^M}\, \varphi = (-{\bf D}-d/2) \varphi =
w \varphi\, , \nn\ee it follows that $\varphi$ is a section of ${\cal
  E}[w]$, {\it i.e.}, a weight $w$ scalar field on $M$.  Moreover,
since the ambient operators~\eqn{fgens} are homogeneous and
tangential, they descend to well-defined operators acting on weighted
tractor bundles over $M$, as listed in~\eqn{operators}. We refer
to~\cite{GP} for details where it is also shown, in an obvious way,
that the same property extends to arbitrary tractor tensors.

Therefore, in the conformally flat context, we have established that the tractor operators are
the generators of the dual ambient conformal group ${\frak{so}}(d+2,2)$ in a momentum representation.
This dictionary between geometry and physics is summarized in Table~\ref{dictionary}.

\begin{figure}
\begin{center}
\begin{tabular}{|c|c|c|c|}
\hline
\multicolumn{2}{c}{\hspace{-.22cm}\vline\hfill Physics}\hfill\vline&\multicolumn{2}{c}{\hfill Mathematics}\hfill\vline\\\hline
Operator&Symbol&Operator&Symbol\\\hline
Momentum&${\bf P}_M$&Canonical Tractor&$X_M$\\
Dilations&${\bf D}$&Weight&$-w-d/2$\\
Rotations&${\bf M}_{MN}$&Double--$D$--Operator& $D_{MN}$\\
Conformal Boosts&${\bf K}_M$&Thomas $D$--Operator&$D_M$\\\hline
\end{tabular}
\end{center}
\caption{\label{dictionary}}
\end{figure}
Next we explain how to read the minimal representation theoretic
relations~\eqn{seven} as relations on tractor identities.
Firstly, the canonical tractor~$X^M$ and Thomas operator $D^M$ are both null. Let us suppose these operators are acting 
on functions of conformal weight ${w}$. Therefore ${\bf D}$ acts as multiplication by $-w-\frac d2$ and
the Thomas operator becomes
\be
D_M=(d+2{w}-2)\partial_M-X_M\partial_N\partial^N\, ,
\nn\ee
in agreement with the results of~\cite{GP}.
Moreover it  follows from~\eqn{seven} that the top slot of the
Thomas operator is given by
\be
{X}_M {D}^M={w}(d+2{w}-2)\, ,
\nn\ee
which agrees with~\eqn{operators}.
From the third identity in~\eqn{seven} we recover the formula relating the product of double-$D$ operators to the
Thomas operator itself~\cite{Gsrni,G}
\be
{D}_{(M}{}^R D_{|R|N)_0}={X}_{(M}{D}_{N)_0}\, ,
\nn\ee
where $(\cdots)_0$ denotes trace-free symmetrization with unit weight.
Other simple yet useful identities, which all follow from~\eqn{seven} include
\be
{X}^N{D}_{MN}={w} X_M\, ,\qquad
{D}_{MN}D^N=-(w-1)D_M\, .
\nn\ee
Moreover, it should be kept in mind that the operators $\{ {\bf
P_M,D,M_{MN},K_M } \}$ only obey the ${\frak{so}}(d+2,2)$ Lie algebra~\eqn{lie
  algebra} when the ambient space is flat. For curved ambient spaces
and therefore general conformal manifolds, this is no longer the
case. This is the subject of the next Section.

\section{Curved Ambient Spaces}

\label{curved}

In this Section we sketch the curved ambient construction of the tractor operators~\cite{GP,CG}
generalizing those above.
We will then show that although a non-vanishing ambient curvature deforms the ${\frak{so}}(d+2,2)$
algebra obeyed in the flat ambient case, the quadratic relations~\eqn{seven} of the minimal representation
persist.

A conformal structure on $M$ determines an FG ambient metric which admits a hypersurface orthogonal
homothety. In the flat case this homothety is generated by the Euler vector field. In the curved ambient construction,
the corresponding homothetic vector field will be denoted by $X$ and its components $X^M$ (which are {\it not}
generally coordinates).
The key identity is then the equation
\be
h_{MN}=\nabla_M X_N\, ,
\nn\ee
where $h$ is the ambient metric and $\nabla$ is its Levi-Civita covariant derivative.
This implies the homothetic conformal Killing equation
\be
{\pounds}_X h = 2 h\, .
\nn\ee
It follows that the one-form dual to $X$ is closed
\be
\nabla_{[M} X_{N]}=0\, ,
\nn\ee
and in fact exact
\be
X_M=\frac12 \nabla_M X^2\,   .
\nn\ee
Clearly, the ambient metric is the double gradient of the homothetic
potential $H=\frac12 X^2$,
\be
h_{MN}=\nabla_M \partial_N H\, .
\nn\ee
The zero locus of the potential $H$ defines the cone $Q$, a quotient of which recovers  
the conformal manifold $M$.
Observe that the above identities for the ambient metric $h_{MN}$ imply that
\be
X^M R_{MNRS}=0=(X^T \nabla_T + 2) R_{MN}{}^R{}_S\, .\label{Xriem}
\ee
To establish a Rosetta Stone between ambient space operators and the
tractor operators~\eqn{operators} on the conformal manifold, our
prescription is to first replace partial derivatives by covariant ones
and coordinates by the components of the homothety $X$ (and then use \cite{CG}).  This gives
the following dictionary for curved ambient generalizations of the
conformal generators
$$
{\bf M}_{MN}\mapsto D_{MN}\equiv 2\ X_{[M}\nabla_{N]}\, ,\qquad\qquad\qquad\qquad\qquad\qquad
$$
$$
{\bf P}_M \mapsto X_M\, ,\qquad\qquad\qquad\qquad
{\bf K}_M\mapsto D_M\equiv2\, (X^N\nabla_N+d/2)\ \nabla_M-X_M\ \Delta\, ,
$$
\be
{\bf D}\mapsto-X^N\nabla_N-d/2\, .\qquad\qquad\qquad\qquad\qquad\qquad\quad
\label{confgens}
\ee
Here, just as in~\eqn{operators}, $X_M$ acts as a multiplication operator.
Importantly these operators act on sections of the tensor bundle on $\wt M$.

These operators are rather remarkable in the sense that the quadratic relations~\eqn{seven}
imposed on ${\cal U}({\frak{so}}(d+2,2))$ by the minimal representation are still valid 
even though they do not, in general, obey the
 ${\frak{so}}(d+2,2)$ dual ambient
conformal algebra.
To see this, we first note that by virtue of the definition of the ambient metric, 
the symbols $X_M$, $\nabla_M$ and $h_{MN}$
obey the same algebra as their flat counterparts $X_M$ (viewed as
coordinates), $\partial_M$ and~$\eta_{MN}$
\be
[\nabla_M,X_N]=h_{MN}\, ,
\nn
\ee
 except that covariant
derivatives no longer commute
\be
\nabla_{[M}\nabla_{N]}=\frac12 R_{MN}^{\, \ts \sharp}\, .
\nn
\ee
The multiplicative operator $R_{MN}^{\, \ts \sharp}$ 
denotes the usual matrix
action of the Riemann tensor (for example $R_{MN}^{\, \ts \sharp}
V^R=R_{MN}{}^R{}_S V^S$). 

Now, using the identity $X^M R_{MN}^{\, \ts \sharp}=0$, it is not
difficult to verify that all the quadratic relations in~\eqn{seven}
hold modulo $X^2$. Moreover the quadratic Casimir continues to
obey~\eqn{c2}. That leaves us only deformations of the 
Lie algebra itself, the non-vanishing ones are
$$
[D_{MN},D_{RS}]-4\ h_{[N[R}D_{M]S]}=4\ X_{[R}X_{[M}R_{N]S]}^{\, \ts
\sharp}\, , $$
$$
[D_{MN},D_R]-2h_{R[N}D_{M]}=2X_R R_{MN}^{\, \ts\sharp}
                             +2X_{[M}R_{N]R}^{\, \ts\sharp}(d+2X.\nabla-2)
                             $$ $$\hspace{3cm}
                             -2X_{R}X_{[M}(R_{N]S}^{\,\ts\sharp}\nabla^S
                                           +\nabla^SR_{N]S}^{\,\ts\sharp})
\, ,$$\vspace{.1cm}
\be
[D_M,D_N]=
\Big(R_{MN}^{\, \ts \sharp}(d+2X.\nabla-4)
+2X_{[M} \ (\nabla^R R_{N]R}^{\, \ts \sharp}
+R_{N]R}^{\, \ts \sharp}\nabla^R)\Big) \ (d+2X.\nabla-2)\, .
\nn\ee
\be
\label{superids}
\ee
(Here the right hand sides should be read as operator expressions and
$X.\nabla\equiv X^N\nabla_N$.) Again, these relations are computed
modulo $X^2$, while tensors of a given weight $ w$ on $M$ correspond
to eigenfunctions of $X.\nabla$ with eigenvalue~$w$.
In particular, note that Thomas D-operators commute on objects
with weight $ w=1-d/2$, in accordance with its interpretation as the
Yamabe operator in that case (see~\eqn{operators}).

Since the left hand sides of~\eqn{superids} are tangential, so too are the right hand sides,
although this is not manifest in the form given above. To manifest tangentiality consider, for example,
the third formula (which was previously given in~\cite{GP} and~\cite{GS}) and firstly replace $X. \nabla$
with $w$ acting on weight $w$ tensors. Moreover, using~\eqn{confgens} and~\eqn{Xriem} it follows that
$(d+2w-2) R^{\ts \sharp}_{MN}\nabla^N=R^{\ts \sharp}_{MN}D^N$. Furthermore, the FG ambient
metric is formally  Ricci flat in any odd dimension (to all orders), and Ricci flat to finite order
 in the defining function~$H$ in even dimensions greater than four. Orchestrating these facts we obtain the simpler
rewriting of the commutator of Thomas-$D$ operators
\be
[D_M,D_N]=(d+2w-4) R_{MN}^{\ts \sharp} + 4 X_{[M} R^{\ts \sharp}_{N]R} D^R\, .
\ee
This identity is valid in dimensions other than four and is fundamental in the sense that the other tractor commutators can be derived from it.

To summarize, we have recovered the usual curved tractor algebra
along with a new interpretation in terms of the ambient conformal
algebra (and its deformation). In particular the ${\frak{so}}(d+2,2)$
generator-valued matrix
\be
J=\left(\begin{array}{ccc}
     w+d/2&-\frac1{\sqrt2}\ X_M&0\\[3mm]
    \frac1{\sqrt2}\ D_N&D_{MN}&
                   \frac1{\sqrt2}\ X_N\\[3mm]
    0&-\frac1{\sqrt2}\ D_M&- w-d/2
        \end{array}\right)\, ,
\nn\ee
obeys $$(J+1)(J+d/2)=0\quad {\rm mod} H\ \quad  \mbox{ and }\ \  {\rm tr} J^2=-\frac12\ d(d+4)\, .$$
This implies that even though the ${\frak{so}}(d+2,2)$ algebra is deformed, the defining
relations~\eqn{seven} of the minimal representation are maintained.

\section{Conclusions}

In this Article we have established a relationship between ambient
tractor operators and symmetries of a physical quantum field
theory. There are many possible implications.  Firstly, in another
direction, tractors in their application to Weyl invariance can be
used to formulate massive and massless theories within a single Weyl
invariant framework~\cite{Gover:2008sw,Gover:2008pt}. From this we
find that Weyl invariance as a fundamental principle for constructing
physical theories yields interesting consequences. In particular,
masses turn out to be tractor weights and very general
Breitenlohner--Freedman stability
bounds~\cite{Breitenlohner:1982jf,Mezincescu:1984ev} follow
directly. The fusion of that work with the direction of this Article
should shed further light on the meaning of mass. In the broader picture
our results here point at a relationship between
conformal field theories and theories in two higher dimensions,
albeit in a dual Fourier transformed space. It is natural to
speculate whether such a relationship may capture aspects of the
AdS/CFT correspondence~\cite{Maldacena:1997re} as a special case.

From the physical standpoint, our model describes new structure on the
geometry of the moduli space of physical excitations for scalar
fields. Since the physical lightcone and its accompanying ambient
space have a natural curved generalization, one wonders whether this
may describe interactions. In particular, renormalization will
generate additional momentum dependence of the effective action for an
interacting scalar field. From this point of view, therefore, a curved
momentum space is most interesting.  A study of the map between curved
ambient spaces and their underlying configuration space theories is
underway~\cite{SW}. In fact theories with a curved momentum space have
been considered before, an example is Snyder's non-commutative, yet
Lorentz covariant geometry which is based upon a de Sitter momentum
space~\cite{S}. Interestingly a ``Snyderspace'', the
supersymmetrization of Snyder's model also exists~\cite{MS}, which
hints that a ``supertractor'' theory may be important.  Since the
gauging of space time algebras technology of~\cite{PvN} also extends
to superalgebras, such a theory certainly exists.

There are many obvious extensions and mathematical applications of the
results obtained. The tractor operators (\ref{fgens}) form the basic
tools for the theory of conformal invariants developed in \cite{G};
that reference is concerned with constructing conformal invariants of
a general conformal manifold. While in low dimensions this gives a
near complete answer, significant gaps remain in high even dimensions.
The results in this Article suggest the exciting possibility that, for
dimension $d$ conformal geometry, the invariant theory might in fact be
best described via ${\frak{so}}(d+2,2)$ representation theory.  

Another interesting observation, is that generic weight tractors can
be extended harmonically away from the cone. From the dual ambient
picture, harmonic objects correspond to those with support on the dual
ambient lightcone. Aside from being a useful computational trick, for
ambient harmonic tractors the Fourier transform yields a symmetric
set-up. Informally, (and using delta distributions as integral kernel
operators) this is depicted by the correspondences
 $$
 \ker \Delta_Y \leftrightarrow \operatorname{coker} X^2 
 $$
 $$
 \operatorname{coker} Y^2 \leftrightarrow \ker \Delta_{X}\, .
$$ This seems particularly interesting when one remembers that an
 ambient harmonic extension corresponds to a collar extension off the
 boundary of a Poincar\'e--Einstein manifold, which is obviously
 relevant to the AdS/CFT correspondence.

On the side of
extensions, an obvious direction is to exploit the ideas here for other
structures. However since a rich picture has emerged from
(essentially) the simplest field equations it seems a first step will
be to explore the ambient interpretation of other equations on the
dual ambient space.

\section*{Acknowledgments}
ARG is supported by Marsden Grant no.\ 06-UOA-029.
A.W. is indebted to the University of Auckland for its warm hospitality.

\end{document}